\documentclass[prd,aps,preprint,tightenlines,showpacs,nofootinbib,superscriptaddress]{revtex4-1}
\usepackage{mathrsfs}
\usepackage{amsfonts}
\usepackage{amsmath}
\usepackage{array}
\usepackage{verbatim}
\usepackage{bm}
\usepackage{epsfig}
\usepackage{graphicx,color}
\usepackage{relsize}
\usepackage{lineno}
\usepackage{multirow}

\RequirePackage{xspace}

\begin{document}

\title{ Probing charmonium-like XYZ states in hadron-hadron ultraperipheral collisions and electron-proton scattering  }

\author{Ya-Ping Xie}
\email{xieyaping@impcas.ac.cn}
\affiliation{Institute of Modern Physics, Chinese Academy of Sciences,
Lanzhou 730000, China}
\author{Xiao-Yun Wang}
\email{xywang@lut.edu.cn}
\affiliation{Department of physics, Lanzhou University of Technology,
	Lanzhou 730050, China}
\author{Xurong Chen}
\email{xchen@impcas.ac.cn}
\affiliation{Institute of Modern Physics, Chinese Academy of
	Sciences, Lanzhou 730000, China}
\affiliation{ University of Chinese
	Academy of Sciences, Beijing 100049, China}
\affiliation{ Institute of
	Quantum Matter, South China Normal University, Guangzhou 510006, China}

\begin{abstract}
Exclusive production of charmonium-like XYZ states in hadron-hadron ultraperipheral collisions(UPCs) and electron-proton scattering is studied employing effective Lagrangian method. Total cross sections and rapidity distributions of charmonium-like XYZ states are obtained in hadron-hadron UPCs and electron-proton scattering process. These predictions can be applied to estimate the observed event number of exclusive charmonium-like XYZ states in hadron-hadron UPCs and electron-proton scattering. The results indicate that it is significant to search $X(3872)$ and $Z^+_c(3900)$ in pA UPCs and Electron-Ion Collider in China will be an advantage platform to observe XYZ states in the future. 
\end{abstract}

\maketitle

\section{Introduction}
The charmonium-like XYZ exotic states have been abundant with experimental progress at BarBar, Belle, BES and LHCb \cite{Brambilla:2010cs,Chen:2016qju,Guo:2017jvc}.  $X(3872)$ was observed in 2003 in $B\to K \pi^+\pi^-J/\psi$ process at Belle experiment \cite{Choi:2003ue}. It was confirmed by CDF, BarBar and D0\cite{Acosta:2003zx, Aubert:2004ns,Abazov:2004kp}. $X(3872)$ was the first example of exotic states because it can't be explained by classical quark model. Since the discovery of $X(3872)$, a couple of other exotic states were discovered near open-charm threshold. In 2005, $Y(4260)$ was discovered in $e^+e^-\to \gamma_{ISR} \pi^+\pi^- J/\psi$ channel at BarBar\cite{Aubert:2005rm}. In 2013, BESIII discovered $Z_c(3900)$ in $J/\psi \pi^\pm$ invariant mass spectrum via $e^+e^- \to J/\psi \pi^\pm \pi^\mp$ \cite{Ablikim:2013mio}. $Z_c(3900)$ was also observed by Belle\cite{Liu:2013dau}.
The CMS measured the $X(3872)$ cross sections at pp collisions at LHC in 2013\cite{Chatrchyan:2013cld}. In 2020, the CMS have 
 observed $B_s^0\to X(3872)\phi$ at pp collisions\cite{Sirunyan:2020qir}.

Although the first exotic state $X(3872)$ has been discovered for 18 years, its nature is elusive now. There are several theoretical models
to describe $X(3972)$ internal structure, such as molecular model\cite{Swanson:2003tb,Wong:2003xk}, tetraquark model \cite{Maiani:2004vq,Ebert:2005nc}. However, there is no determination of nature for $X(3872)$. The situations are same to $Y(4260)$ and $Z_c(3900)$. 

In the next decade, the Electron-Ion Collider (EIC) will be built for studying the structure of nucleon and exotic states\cite{Accardi:2012qut}, for example, EIC-US\cite{eic-us} and EIC in China (EicC)\cite{Chen:2020ijn,eicc}. 
Investigation of nature of exotic states is important in proposed EICs. As a consequence, it is necessary to predict the photoproduction of XYZ states production for proposed EICs. These predictions will be helpful for the design of EICs.

 On theoretical side, photoproduction of exotic states is investigated in several models adopting effective Lagrangian method. 
 For example, photoproduction of $Z_c(4430)$, $Z_c(4200)$ and $Z_c(3900)$
 were studied in Refs.\cite{Liu:2008qx,Wang:2015lwa,Galata:2011bi,Lin:2013mka, Klein:2019avl}. At the same time, photoproduction of pentaquark states has been investigated in Refs.\cite{Winney:2019edt,Goncalves:2019vvo,Cao:2019gqo,Xie:2020ckr,Xie:2020niw,Xie:2020wfe}.
In Ref.\cite{Albaladejo:2020tzt}, the authors have studied photoproduction of several charmonium-like XYZ states in photon-proton scattering applying effective Lagrangian method. It is necessary to predict the production of charmonium-like XYZ states in electron-proton scattering for proposed EICs. Thus, we extend the production of XYZ states to electron-proton scattering in this manuscript. 

In hadron-hadron collisions, when the sum of impact parameters of two hadrons is large, strong interaction between hadrons can be neglected since strong interaction is short range interaction. However, the electromagnetic interaction between two hadrons is important in this situation. This collision is referred to ultraperipheral collision (UPC)\cite{upc}. UPCs are important experiments to study the structure of hadron and photoproduction of vector mesons and exotic states. In the hadron-hadron UPCs, the virtuality of the photon is $-q^2 <(1/R_A)^2$ which is very small in large $R_A$  \cite{Bertulani:2005ru}. Therefore, the photon can be viewed as a real photon. 

In electron-proton scattering, the electron interacts with proton via photon. The total cross section of the electron-proton can be also cast into photon flux and photon-proton cross section. in electron-proton scattering, the virtuality of photon is $-q^2 = 4EE^\prime\sin(\theta/2)$, which is not small. Therefore, the photon in electron-proton scattering can be treated as virtual photon. 

In the chamonium-like XYZ production, the final states include charmonium states. Thus, the charmonium states production is also important in EICs predictions. On the other hand, the near threshold charmonium production can be employed to investigate proton mass decomposition\cite{Wang:2019mza,Xie:2021seh}. Currently, the GlueX Collaboration measured $J/\psi$ production near threshold using proton-fixed target \cite{Ali:2019lzf}. 

The aim of this manuscript is to predict exclusive photoproduction of charmonium-like XYZ in hadron-hadron UPCs at the RHIC and LHC and electron-proton scattering for proposed EICs. These predictions can help experimental collaborations to estimate the charmonium-like XYZ production in hadron-hadron UPCs and electron-proton scattering for proposed EICs. 

This paper is organized as follows. Theoretical framework will be presented in Sec.II. Numerical results and discussions will be exhibited in Sec. III. Conclusions will be given at Sec. IV. 
\section{Theoretical framework}
In hadron-hadron UPCs and electron-proton scattering, the total cross section can be factorized into photon flux and cross section of photon-proton scattering. The photon flux of hadron and electron can be obtained from phenomenological models. The cross section in photon-proton scattering can be obtained in the effective Lagrangian method. The cross section of $\sigma^{\gamma p\to Xn}(W)$ is integrated by $|t|$
\begin{equation}
\sigma^{\gamma p\to Xn}(s)=\int dt\frac{d\sigma^{\gamma p\to Xn}}{dt}(s,t).
\end{equation}
where $X$ denotes charmonium-like XYZ exotic states and $n$ denotes proton and neutron.
The differential cross section of $\gamma p\to Xn$ is obtained as\cite{Liu:2008qx}
\begin{eqnarray}
\frac{d\sigma^{\gamma p\to Xn}}{dt}(s,t)=\frac{1}{64\pi s |k_{1cm}|^2}
|\frac{1}{4}\mathcal{M}^{\gamma p\to Xn}(s,t)|^2,
\label{dsigma1}
\end{eqnarray}
when the amplitude square is summed up of helicities as
\begin{eqnarray}
|\mathcal{M}(s,t)|^2 = \sum_{pol}|\langle \lambda_X \lambda_N^\prime|\mathcal{T}|\lambda_{\gamma}\lambda_N\rangle|^2.
\end{eqnarray}
In this manuscript, we follow the scattering amplitude of XYZ states as same as calculations in JPAC's paper\cite{Albaladejo:2020tzt} and references therein. In this manuscript, we only represent the basic expressions of the scattering amplitudes.
In the process of $\gamma p\to Y(4260)p$, when the energy is in high limit, the amplitude can be written as \cite{Albaladejo:2020tzt} 
\begin{eqnarray}
\langle \lambda_Y \lambda_N^\prime|\mathcal{T}|\lambda_{\gamma}\lambda_N\rangle
=F(s,t)\delta_{\lambda_{\gamma},\lambda_Y}\delta_{\lambda_N,\lambda_N^{\prime}}.
\end{eqnarray}
While in the low energy limit, the scattering amplitude is given as \cite{Winney:2019edt}
\begin{eqnarray}
\langle \lambda_Y \lambda_N^\prime|\mathcal{T}|\lambda_{\gamma}\lambda_N\rangle
=\frac{F(s,t)}{s}[\bar{u}(p^\prime,\lambda_N^\prime)\gamma_\mu u(p,\lambda_N)]\notag\\
\times \epsilon_\nu^*(q^\prime, \lambda_Y)[\epsilon^\mu(q,\lambda_\gamma)q^\nu-\epsilon^\nu(q, \lambda_{\gamma})q^\mu].
\end{eqnarray}
$F(s,t)$ is given as \cite{Albaladejo:2020tzt}
\begin{equation}
	F(s,t) = ieA_\psi\Big(\frac{s-s_{th}}{s}\Big)^{a(t)}e^{bt}.
\end{equation}
The scattering matrix of $\gamma p\to Zn$ reads \cite{Lin:2013mka}
\begin{eqnarray}
\langle \lambda_Z \lambda_N^\prime|\mathcal{T}|\lambda_{\gamma}\lambda_N\rangle=
\frac{g_{VZ\pi}}{m_Z}\epsilon_\mu(q, \lambda_V)
	\epsilon_\nu^*(q^\prime, \lambda_Z)[(q\cdot k)g^{\mu\nu}-k^\mu q^\nu]\notag\\
	\times \sqrt{2} g_{\pi NN}\beta(t)\bar{u}(p^\prime,\lambda_N^\prime)\gamma^5u(p,\lambda_N).
\end{eqnarray}
Two models for $\beta(t)$ are introduced. One is fixed-spin model (low energy limit) and the other one is Regge  model (high energy limit)\cite{Albaladejo:2020tzt}. 

For X(3872), the scattering matrix of $\gamma p \to X(3872)p$ can be expressed as \cite{Hanhart:2011tn}
\begin{eqnarray}
\langle \lambda_X \lambda_N^\prime|\mathcal{T}|\lambda_{\gamma}\lambda_N\rangle=g_{\psi X \varepsilon}\epsilon_{\sigma\nu\alpha\beta}[g^{\sigma\mu}\epsilon^{*\nu}(q^\prime,\lambda_X)q^\alpha\epsilon^{\beta}(q,\lambda_\varepsilon)]\notag\\
\times \beta(t)\bar{u}(p^\prime,\lambda_N^\prime)\big(g_{\epsilon NN }\gamma^\mu + g_{\epsilon NN}^\prime 
\frac{\sigma^{\mu \nu}k_\nu}{2m_N}\big)u(p,\lambda_N).
\end{eqnarray}
The coupling constants are determined by decay processes. For $\beta(t)$, fixed-spin model and Regge model are introduced as same as $Z_c^+(3900)$. 

In this manuscript, we focus on the production of charmonium-like XYZ in hadron-hadron UPCs and electron-proton scattering. The rapidity distribution of XYZ states in UPCs is the product of photon flux and cross sections of $\gamma p\to Xn$.  The rapidity distribution of XYZ particles in proton-proton UPCs is given as follows \cite{Coelho:2020lyd}
\begin{eqnarray}
\frac{d\sigma^{pp\to Xpp}}{d\mathrm{y}}=k^+\frac{dN_\gamma (k^+)}{dk^+}\sigma^{\gamma p\to Xn}(W^+)+k^-\frac{dN_\gamma (k^-)}{dk^-}\sigma^{\gamma p\to Xn}(W^-).
\label{dsdy}
\end{eqnarray}
In above equation, $k$ is momentum of the radiated photon from proton. $\mathrm{y}$ is the rapidity of the exotic meson.  $k^{\pm}=M_V/2\exp(\pm |\mathrm{y}|)$.  $\mathrm{W}^{\pm}$ is the center mass energy of photon-proton system In UPCs, $W^{\pm}=(2k^\pm\sqrt{s})^{1/2}$ with $\sqrt{s}$ center-energy. $dN_\gamma (k)/dk$ is photon flux. It is given as \cite{Bertulani:2005ru, Xie:2018rog}
\begin{eqnarray}
\frac{dN_{\gamma}(k)}{dk}=\frac{\alpha_{em}}{2\pi k}\Big[1+\Big(1-\frac{2k}{\sqrt{s}}\Big)^2\Big]\Big(
\ln \Omega-\frac{11}{6}+\frac{3}{\Omega}-\frac{3}{2\Omega^2}+\frac{1}{3\Omega^3}\Big),
\end{eqnarray}
where $\alpha_{em}$ is the fine structure constant and $\Omega = 1+0.71 \mathrm{GeV}^2/Q^2_{min}$, with $Q^2_{min} = k^2/\gamma_L^2$, $\gamma_L$ is the Lorentz boost factor of the hadron with
$\gamma_L = \sqrt{s}/2m_p$.

In $pA$ UPCs, the charge number of nucleus is much larger than the proton. Due to the photon flux is proportional to
the square of the charge number Z, the photon flux of proton is much smaller than the nucleus. Therefore, the contribution of photon from
proton can be neglected in pA UPCs. The production of XYZ exotic states in $\mathrm{p}\mathrm{A}$ UPCs can be written as 
\begin{eqnarray}
\frac{d\sigma^{pA\to XpA}}{d\mathrm{y}}=k\frac{dN_{\gamma }(k)}{dk}\sigma^{\gamma
	p\rightarrow Xn}(W).
\end{eqnarray}
The photon flux emitted from nucleus is given as \cite{Klein:1999qj}
\begin{equation}
\frac{dN_{\gamma }(k)}{dk}=\frac{2Z^{2}\alpha_{em} }{\pi k}\big(XK_{0}(X)K_{1}(X)-%
\frac{X^{2}}{2}[K_{1}^{2}(X)-K_{0}^{2}(X)]\big).\label{phflux}
\end{equation}
where $X=2kR_A/\gamma_L$ and $R_A$ is the radius of nucleus, $K_0(x)$ and $K_1(x)$ are second kind Bessel functions. Adopting the photon flux and photon-proton cross section, one can
obtain the differential cross sections of XYZ exotic states in hadron-hadron UPCs. 

In the electron-proton scattering, the total cross sections of XYZ exotic states can be written as
\cite{Lomnitz:2018juf,Klein:2019avl}
\begin{equation}
\sigma (ep\rightarrow eXn)=\int dkdQ^{2}\frac{dN^{2}(k,Q^{2})}{dkdQ^{2}}%
\sigma _{\gamma ^{\ast }p\rightarrow Xn}(W,Q^{2}),
\end{equation}
where $k$ is the momentum of the photon emitted from initial electron in proton rest
frame. $W$ is the center of mass (c.m.) of the photon-proton system and $Q^{2}$ is the virtuality of the virtual photon. 
The photon flux of electron can be expressed as \cite{Budnev:1974de}
\begin{equation}
\frac{d^{2}N(k,Q^{2})}{dkdQ^{2}}=\frac{\alpha_{em} }{\pi kQ^{2}}\Big[1-\frac{k}{%
	E_{e}}+\frac{k^{2}}{2E_{e}^{2}}-\Big(1-\frac{k}{E_{e}}\Big)\Big|\frac{%
	Q_{min}^{2}}{Q^{2}}\Big|\Big].
\end{equation}
Here $E_e$ is the energy of the initial electron, $k$ is the momentum and $Q^2$ is the virtuality of the virtual photon. 
$Q_{min}^2$ is the minimum of the virtuality and it is given as \cite{Lomnitz:2018juf}
\begin{eqnarray}
Q^2_{min} = \frac{m_e^2k^2}{E_e(E_e-k)},
\end{eqnarray}
where $m_e$ is the mass of electron.

The $Q^{2}$ dependence of $\sigma _{\gamma ^{\ast }p\rightarrow
	Xn}(W,Q^{2})$ is factorized as
\begin{equation}
\sigma _{\gamma ^{\ast }p\rightarrow Xn}(W,Q^{2})=\sigma _{\gamma
	p\rightarrow Xn}(W,Q^{2}=0)\bigg(\frac{M_{V}^{2}}{M_{V}^{2}+Q^{2}}\bigg)%
^{\eta }.
\end{equation}
Here $\eta$ is the scale function of $Q^2$ dependent. $\eta = c_1+c_2(Q^2+M_V^2)$, with $c_1 = 2.36$ and $c_2$ = 0.0029 $\mathrm{GeV}^{-2}$\cite{Lomnitz:2018juf}. The data of $c_1$ and $c_2$ are determined by fitting from HERA data. 

Employing the amplitude of $\gamma p \to Xn$ and photon flux of hadrons, we can obtain the differential cross section in hadron-hadron UPCs and electron-proton scattering.
Then, we can obtain the rapidity distributions and total cross sections of XYZ states in hadron-hadron UPCs and electron-proton scattering. 
\section{Numerical results}
 The production of XYZ exotic states is important in studying the nature of XYZ exotic states. In this work, we calculate the total cross sections of three typical XYZ states ($X(3872)$, $Y(4260)$ and $Z_c^+(3900)$) in hadron-hadron UPCs and electron-proton scattering. These total cross sections can help experimental collaborations to estimate the production of XYZ states before to observe of XYZ states. The calculation details are presented in section II. In this section, we will exhibit the numerical results and give some discussions
 
Firstly, the total cross sections of XYZ states in hadron-hadron UPCs and electron-proton scattering are listed 
 in Table.~\ref{table00} and Table.~\ref{table01}. The cross sections in UPCs are exhibited in Table.~\ref{table00}. The total cross sections of  left small rapidity regions are obtained using low energy models while predictions of right large rapidity regions are calculated employing high energy models.  In pp UPCs calculation, only the high energy models are adopted. 
 
\begin{table}[h]
	\begin{tabular}{c|c|c|c|c|c}
		\hline
		\hline
		  & \multicolumn{2}{c|}{ $\mathrm{p}$-$\mathrm{Pb}$}   &   \multicolumn{2}{c|}{ $\mathrm{p}$-$\mathrm{Ar}$} &   $\mathrm{p}$-$\mathrm{p}$    \\
		\hline
				  & \multicolumn{2}{c|}{$\sqrt{s} = $ 200 GeV}   &   \multicolumn{2}{c|}{ $\sqrt{s} = $ 110 GeV} &  $\sqrt{s} = $ 14 TeV   \\
		\hline
		&$-4 <\mathrm{y} < -1$ &  $-2 <\mathrm{y} < 1$ & $1.5 <\mathrm{y} < 4$& $3 <\mathrm{y} < 6$& $-6 <\mathrm{y} < -3$\\
		\hline
		$X(3872)$ &12 mb &3.5 nb & 0.30 mb & 0.13 nb&4.1 pb\\
		\hline
				$Y(4260)$ &2.0 mb &3.8 mb & 46 nb & 79 nb& 20 nb\\
		\hline
				$Z_c^+(3900)$ &1.7 mb &37 nb & 57 nb & 2.0 nb&  28 pb\\
		\hline
		\hline
	\end{tabular}\caption{ Total cross sections of XYZ  states in hadron-hadron UPCs at the RHIC and LHC. The collider energies are listed in second row and rapidity regions are in third row. The predictions of small rapidities of pPb and pAr UPCs are performed in low energy models while
the predictions in large rapidities are employed in high energy models. The predictions of pp UPCs are adopted in high energy models.}
	\label{table00}
\end{table}
From Table.~\ref{table00}, it can seen that the cross sections of $X(3872)$ and $Z_c^+(3900)$ in low energy models are several mb The numerical results indicate that the cross sections of $X(3872)$ and $Z_c^+(3900)$ are large in $pA$ UPCs at the RHIC and LHC. It is easy to observe the $X(3872)$ and $Z_c^+(3900)$ in pPb and pAr UPCs at RHIC and LHC.
However, in pp UPCs, the cross sections of XYZ are smaller than in pA UPCs since the collider energy of proton-proton is too high. It is difficult to observe XYZ states in pp UPCs at LHC. 

The cross sections of  XYZ states in electron-proton scattering for two proposed EICs are 
displayed in Table.~\ref{table01}, where the cross sections of EicC are obtained using low energy models while predictions of EIC-US are computed in high energy models. In electron-proton scattering, the region of $\mathrm{Q}^2$ is 0.01 $\mathrm{GeV}^2 <\mathrm{Q}^2 <$ 1.0 $ \mathrm{GeV}^2$.

\begin{table}[h]
	\begin{tabular}{c|c|c}
		\hline
		\hline
		&  EicC & EIC-US    \\
				\hline
		&  $\sqrt{s} = $ 16.7 GeV & $\sqrt{s}$ = 140.7 GeV   \\
		\hline
		&   W $< 16$ GeV &  20 GeV $< $ W $<$ 60 GeV   \\
		\hline
		$X(3872)$ &1.2 nb & 0.21 pb\\
		\hline
		$Y(4260)$ &0.20 nb & 2.0 nb\\
		\hline
		$Z_c^+(3900)$ &0.16 nb & 0.48 pb\\
		\hline
		\hline
	\end{tabular}\caption{ Total cross sections of XYZ exotic states in two W regions in electron-proton scattering for proposed EICs, with 0.01 $\mathrm{GeV}^2 <\mathrm{Q}^2 <$ 1.0 $ \mathrm{GeV}^2$. The collider energies are listed at second row and the W regions are in third row.The
    predictions of EicC are performed in low energy models while the predictions of EIC-US are employed in high energy models.}
	\label{table01}
\end{table}

From Table.~\ref{table01}, it can be seen that the low energy models give much larger cross sections than high energy models in EICs for $X(3872)$ and $Z_c^+(3900)$. These can be concluded that it is much easier to observe $X(3872)$ and $Z_c^+(3900)$ states in low W regions for the proposed EICs. 

\begin{figure}[h]
	\centering
	\includegraphics[width=0.48\textwidth]{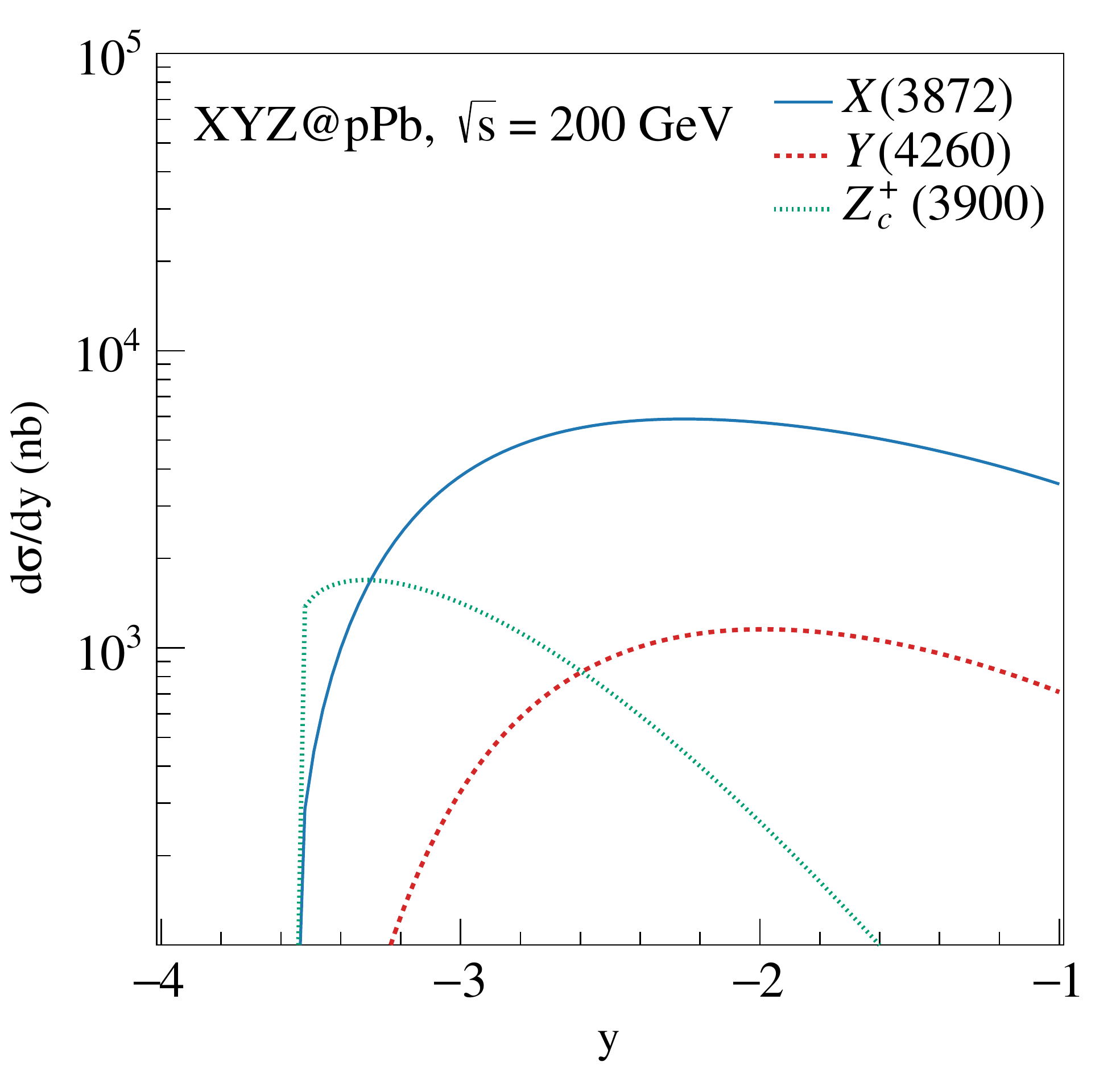}
	\includegraphics[width=0.48\textwidth]{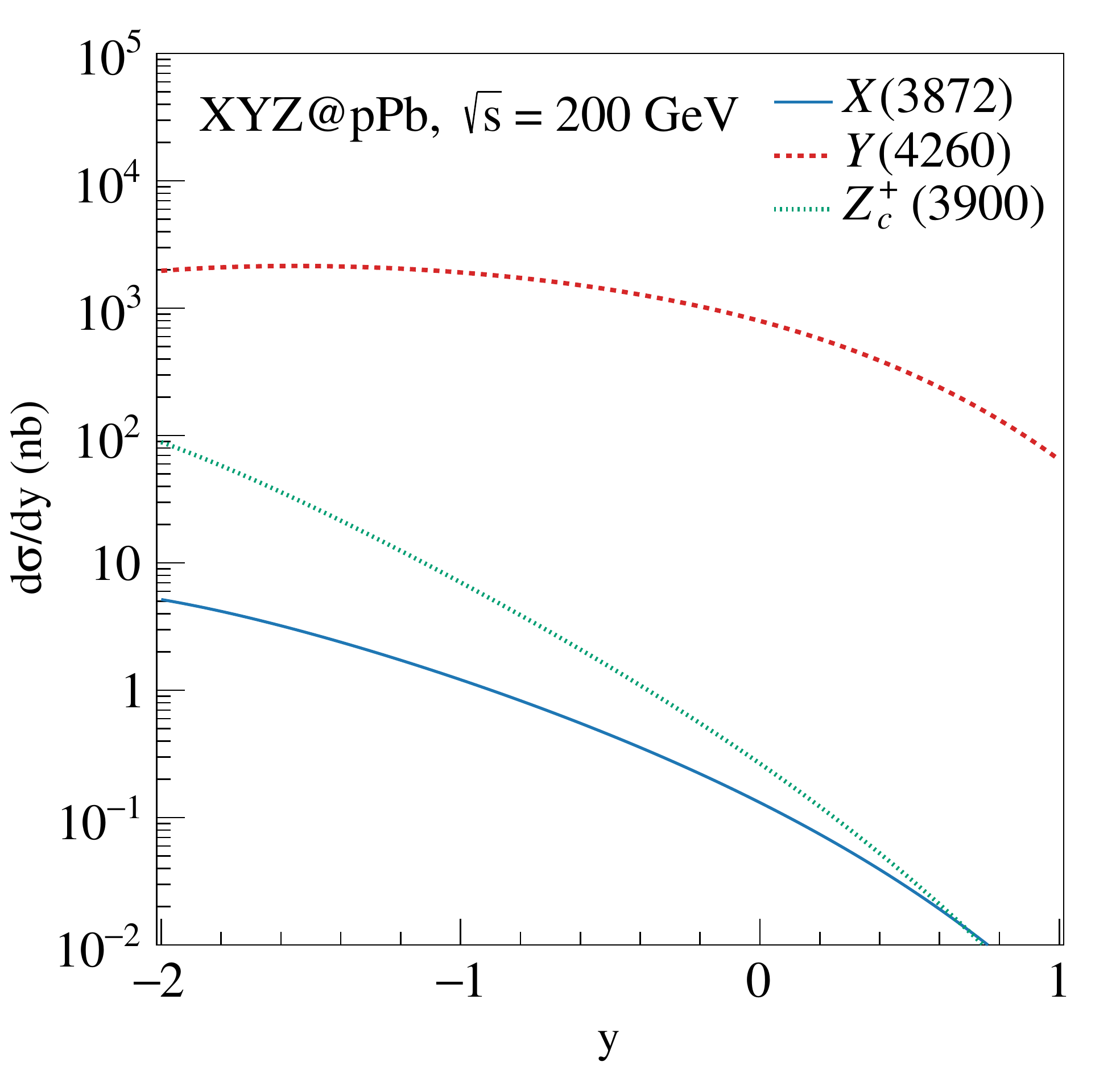}
	\caption{(Color online) Rapidity distributions of XYZ particles in $\mathrm{p}\mathrm{Pb}$ UPCs at $\sqrt{s}$ = 200 GeV at RHIC. The beam of lead is moving in positive rapidity direction. The predictions in left panel are calculated in low energy models while the predictions in the right panel are adopted in high energy models.}
	\label{fig01}
\end{figure}

\begin{figure}[h]
	\centering
	\includegraphics[width=0.48\textwidth]{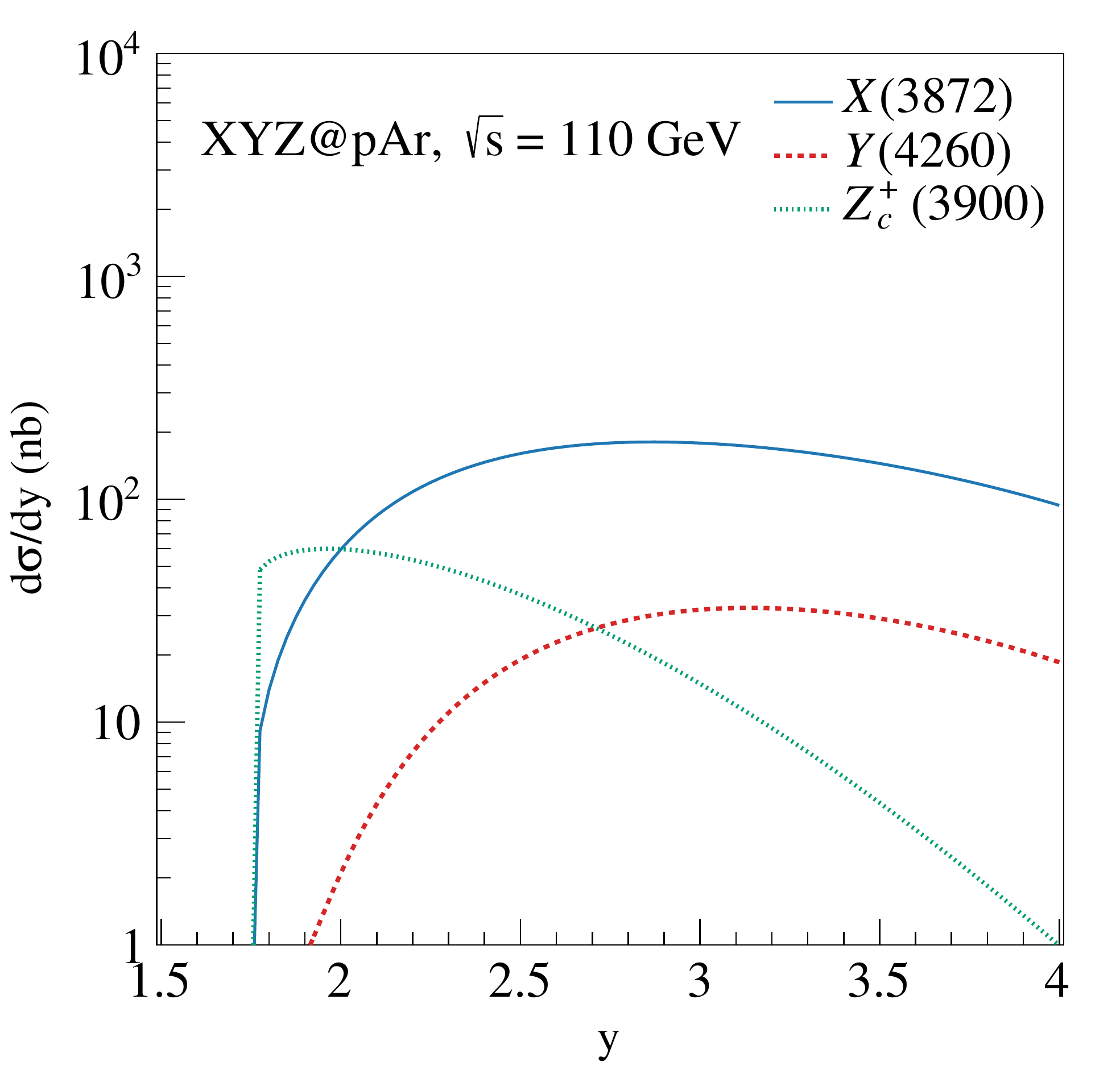}
	\includegraphics[width=0.48\textwidth]{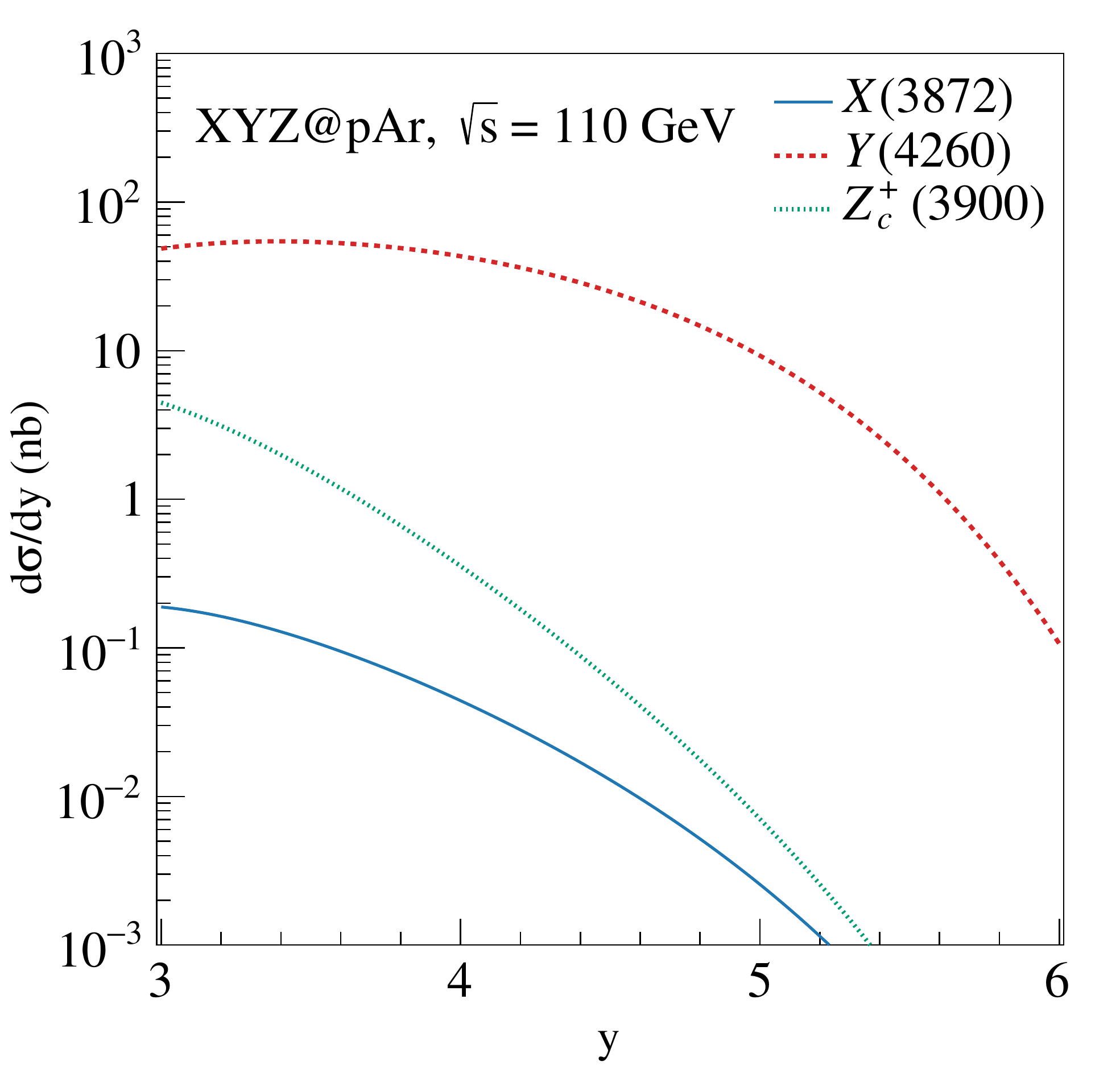}
	\caption{(Color online) Rapidity distributions of $XYZ$ in $\mathrm{p}\mathrm{Ar}$ fixed target system at $\sqrt{s}$ = 110 GeV at LHCb. Proton is the fixed target and beam of Argon is moving in positive rapidity direction. The predictions in left panel are calculated in low energy models while the predictions in the right panel are adopted in high energy models.}
	\label{fig02}
\end{figure}

Secondly, the rapidity distributions of XYZ in UPCs in $\mathrm{p}\mathrm{Pb}$ are shown in Fig.~\ref{fig01}. The predictions of low energy models are depicted in left panel while predictions of high energy models are illustrated in right panel. The positive rapidity is defined as the direction of beam of lead moving. In the graphs, the solid curves are predictions of $X(3872)$, the dashed curves are predictions of $Y(4260)$ and the dotted curves are results of $Z^+_c(3900)$ hereafter. 
The rapidity distributions of XYZ states in $\mathrm{p}\mathrm{Ar}$ fixed target are illustrated in Fig.~\ref{fig02}. The predictions of low energy models are depicted in left panel while predictions of high energy models are described in right panel. The beam of argon moves in positive rapidity direction.Because the proton is fixed in $\mathrm{p}\mathrm{Ar}$ collisions, the XYZ are produced in larger rapidity regions than pPb collisions. 

\begin{figure}[h]
	\centering
	\includegraphics[width=0.48\textwidth]{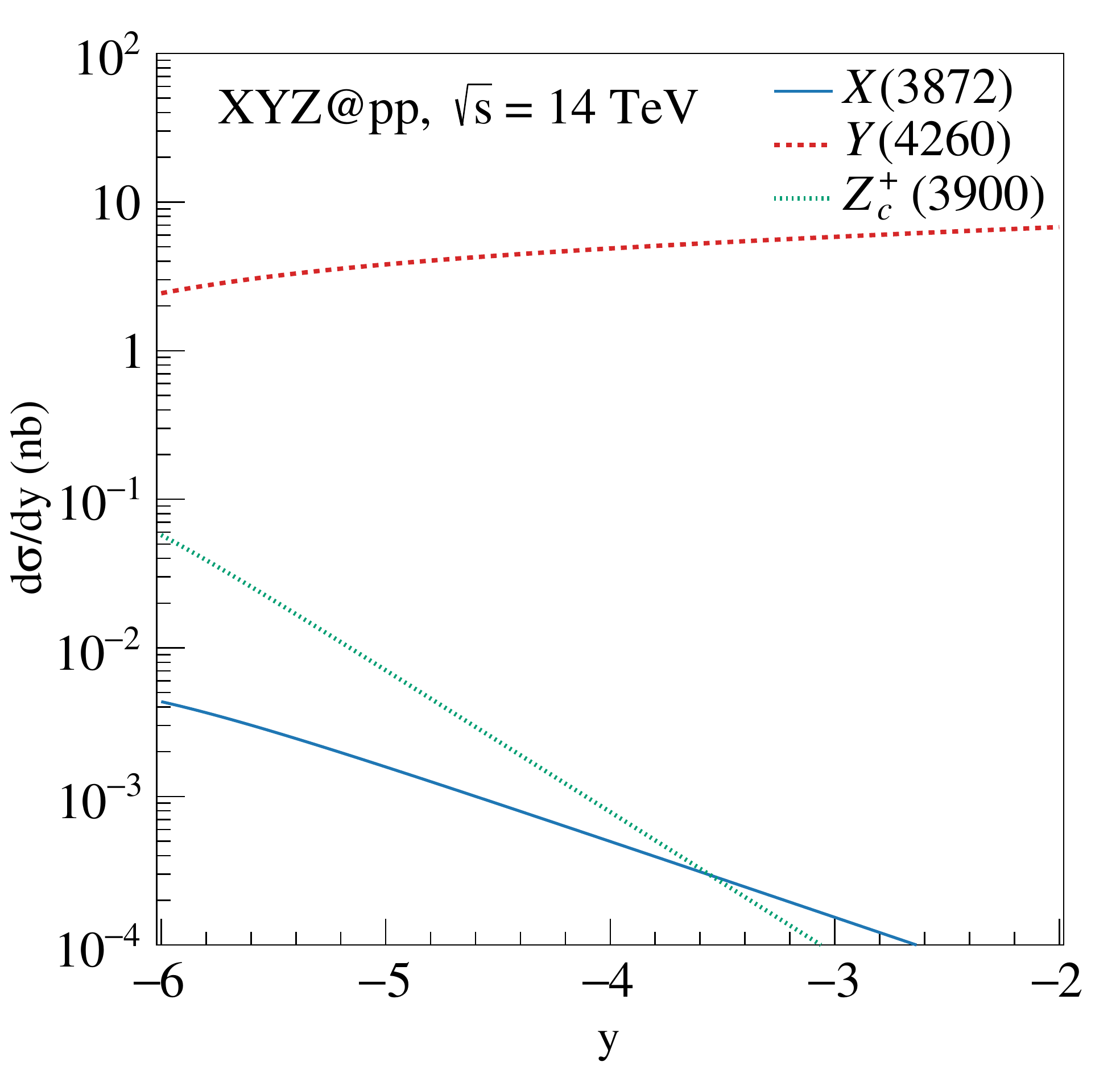}
	\includegraphics[width=0.48\textwidth]{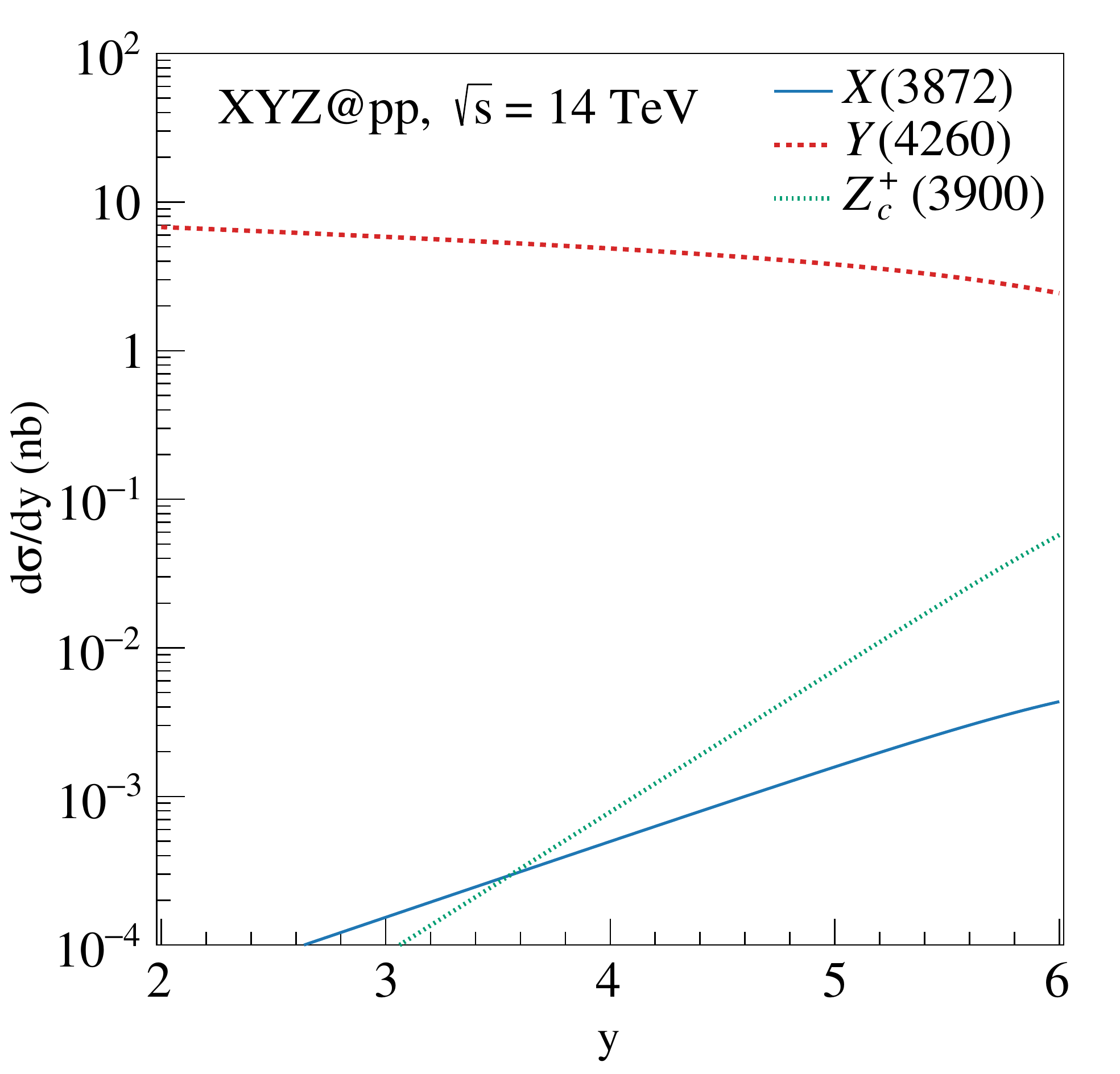}
	\caption{(Color online) Rapidity distributions of $XYZ$ in  pp UPCs at $\sqrt{s}$ = 14 TeV  for LHCb. The predictions in two panel are 
	calculated in high energy models.}
	\label{fig03}
\end{figure}

Next, the predictions of XYZ in proton-proton UPCs are shown in Fig.~\ref{fig03}. In pp UPCs, only the high energy models is adopted as the collider energy in proton-proton UPCs is very high. In pp UPCs, there are two photon sources from proton in each directions. Therefore, there are two produced XYZ in rapidity regions.
The predictions indicate the cross sections of $X(3872)$ and $Z_c^+(3900)$ in pp UPCs are small. However, the cross sections of $Y(4260)$ are large in pp collisions at LHC.

\begin{figure}[h]
	\centering
	\includegraphics[width=0.48\textwidth]{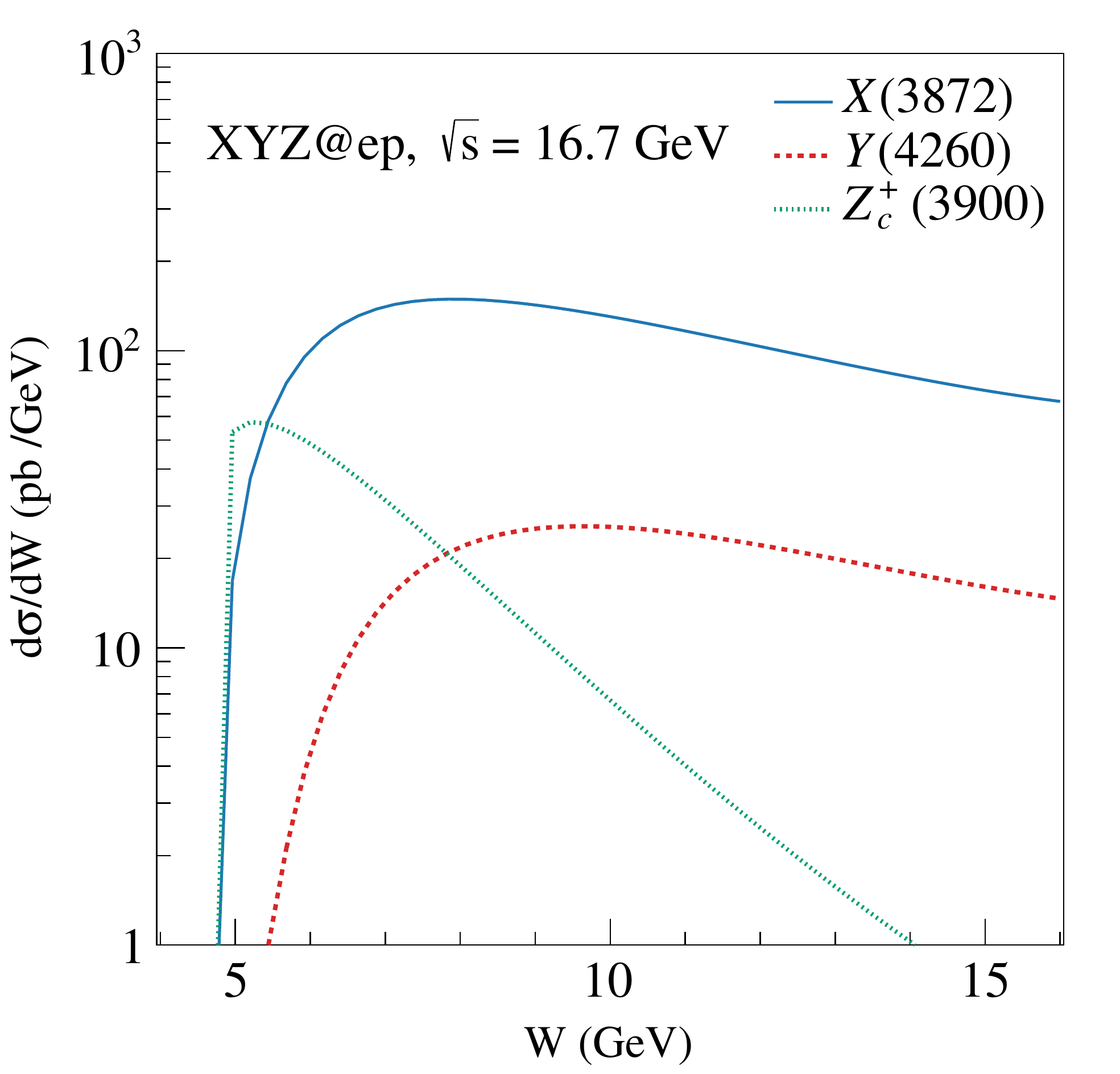}
	\includegraphics[width=0.48\textwidth]{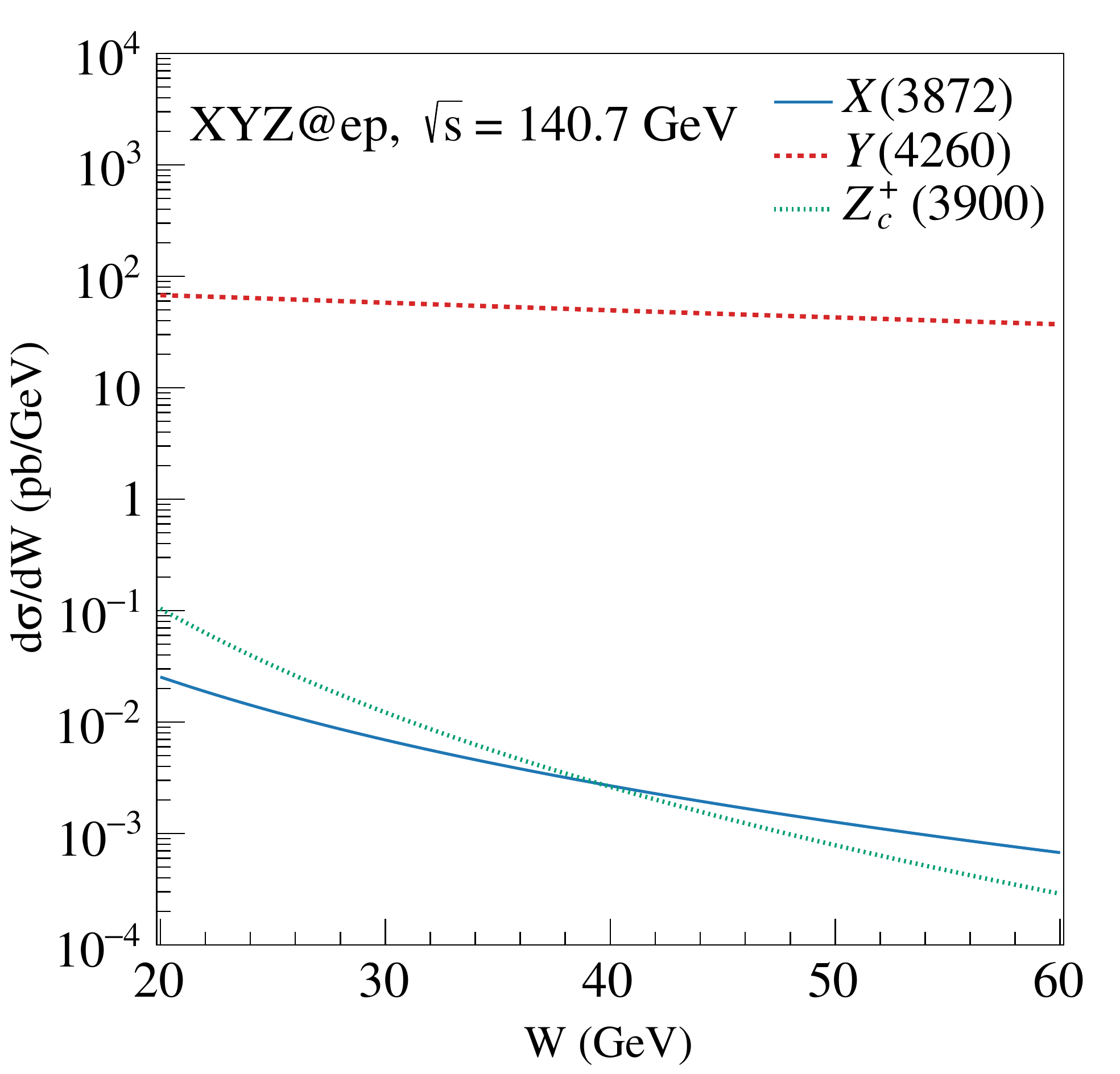}
	\caption{(Color online) Cross section as a function of W for $XYZ$ in electron-proton scattering at EicC (left panel)  and EIC-US (right panel). The region of $\mathrm{Q}^2$ is 0.01 $\mathrm{GeV}^2 < \mathrm{Q}^2 <$ 1.0 $\mathrm{GeV}^2$. The predictions in left panel are calculated in low energy models while the predictions in the right panel are adopted in high energy models.}
	\label{fig04}
\end{figure}

Moreover, we display the predictions of XYZ exotic states in electron-proton scattering for two proposed EICs in Fig.~\ref{fig04}. The region of $\mathrm{Q}^2$ is 0.01 $\mathrm{GeV}^2 < \mathrm{Q}^2 <$ 1.0 $\mathrm{GeV}^2$. The predictions of EicC are described in left panel using low energy models while the predictions of EIC-US are depicted in right panel using high energy models. The results in the two graphs indicate that the XYZ exotic states are mostly produced in low W regions which is the same as photon-proton scattering. Therefore, the GlueX also can observe the XYZ states at JLab. This indicate that EIC-US and EicC can observe XYZ states in low energy regions in the future.

Finally, it can be seen that the cross sections of $X(3872)$ or $Z_c^+(3900)$ states adopting low energy models are much larger than high energy models. The cross sections at $\mathrm{p}\mathrm{Pb}$ or $\mathrm{p}\mathrm{Ar}$ are larger than the cross sections in $\mathrm{p}\mathrm{p}$ UPCs. 
Consequently, it is better to search $X(3872)$ or $Z_c^+(3900)$ states in $\mathrm{p}\mathrm{Pb}$ than in $\mathrm{p}\mathrm{p}$ UPCs. EicC is a ideal platform to study the XYZ exotic states production in the future. 
\section{Conclusion}
In this manuscript, we study the predictions of three typical charmonium-like XYZ states in hadron-hadron UPCs at the RHIC and LHCb and electron-proton scattering for proposed EICs. Effective Lagrangian method is adopted in the calculation of photoproduction of XYZ states. 
The total cross sections are displayed in this paper for UPCs and electron-proton scattering. 
 The rapidity distributions of charmonium-like XYZ mesons are presented in this paper for hadron-hadron UPCs at the RHIC and LHC. The total cross section are also shown for hadrons UPCs and electron-proton scattering. 
 
 The cross sections of XYZ states in hadron-hadron UPCs imply that XYZ states  can be observed at pPb UPCs at RHIC. Thus, we suggest STAR
 collaboration to perform $X(3872)$ or $Z_c^+(3900)$ observation in the future experiments. The production of XYZ states can help us to understand the nature of exotic states and exclude some theoretical models of exotic states. It is of great significance to perform observations of XYZ states. 
 
 For the fixed-target system, the cross sections of XYZ states are large at central rapidity regions. It is not difficult to observe XYZ states at fixed-target system. The predictions in EICs indicate that cross sections of $X(3872)$ and $Z_c^+(3900)$ are large at low energies region. In the high energies region, the cross sections are smaller than the low energies region . Thus, EicC is an advantage platform to observe $X(3872)$, $Y(4260)$ and $Z_c^+(3900)$ states in the future.
 
 We hope that our predictions can help experimental collaborations to estimate the production of charmonium-like XYZ states before to observe the XYZ states, which is variable to investigate the nature of exotic states.
 
\section*{Acknowledgment}
The authors thank the discussions with Dr. D. Winney.
The work is partially supported by Key Research Program of the Chinese Academy of Sciences (Grant NO. XDB34030301) and National Natural Science Foundation of China (Grant No. 12065014). This work is also partly supported by the HongLiu Support Funds for Excellent Youth Talents 
of Lanzhou University of Technology.

\end{document}